\documentclass[12pt]{article}
\usepackage{epsfig}
\usepackage{graphics}
\usepackage{times}
\begin{document}

\title{Is It Small-scale Weak Magnetic Activity That Effectively Heats the Upper Solar Atmosphere?}
\author {K. J. LI$^{1, 3, 4}$, J. C. XU$^{1, 3, 4}$, W.  FENG$^{2}$     \\
\footnotesize{$^{1}$Yunnan Observatories, Chinese Academy of Sciences, Kunming 650011, China}\\
\footnotesize{$^{2}$Research Center of Analysis and Measurement, Kunming University of Science and Technology, Kunming 650093, China}\\
\footnotesize{$^{3}$Center for Astronomical Mega-Science, Chinese Academy of Sciences, Beijing 100012, China} \\
\footnotesize{$^{4}$Key Laboratory of Solar Activity, National Astronomical Observatories, CAS, Beijing 100012, China} \\
}

\date{}
\baselineskip24pt
\maketitle

\begin{abstract}
Solar chromosphere and coronal heating is a big question for astrophysics.
Daily measurement of 985 solar spectral irradiances (SSIs) at the spectral intervals 1-39 nm and 116-2416 nm during March 1 2003 to October 28 2017  is utilized to investigate phase relation respectively with daily sunspot number, the Mount Wilson Sunspot Index, and the Magnetic Plage Strength Index. All SSIs which form in the whole heated region: the upper photosphere, chromosphere, transition region, and corona are  found to be significantly more correlated  to weak magnetic activity than to strong magnetic activity, and to dance in step with  weak magnetic activity.
All SSIs which form in the low photosphere (the unheated region), which indicate the ``energy" leaked from the solar subsurface
are  found to be more related to strong magnetic activity instead and in anti-phase with weak magnetic activity. In the upper photosphere and chromosphere, strong magnetic activity should lead SSI by about a solar  rotation, also displaying that weak magnetic activity should take effect on heating there. It is thus small-scale weak magnetic activity that effectively heats the upper solar atmosphere.\\
{\bf keywords}  Sun: corona -- Sun: activity -- Sun: atmosphere
\end{abstract}

\section{Introduction}
For more than 70 years since the 1940s, it has been a challenging puzzle why the solar corona is much hotter than the underlying chromosphere and photosphere  and how the energy of corona heating is transported upwards and dissipates there (Edlen 1945; De Moortel $\&$ Browning 2015).
Up to now, plenty of advances have been achieved in observation and theory studies for coronal heating, with extraordinary progress especially in the recent decades (Klimchuk 2006, 2015; Parnell $\&$ De Moortel 2012; De Moortel $\&$ Browning 2015). On one hand, recent high-resolution observations display that
ubiquitous small-scale (a dozen arc seconds or less) isolated magnetic elements, such as network and intra-network magnetic fields and ephemeral regions, cover the solar surface like a magnetic blanket (Zirin 1988; Wilhelm et al 2007), and small-scale magnetic activity phenomena, which are mainly related to these small-scale magnetic elements, frequently occur at the solar atmosphere to release energy there. They may be generally divided into the following  groups: (1)spicules and macro-spicules; (2)jets, including surges,  extreme ultraviolet jets, and X-ray jets; (3) bright (dark) point features, e.g. network bright points, X-ray bright points  (size: $\sim 10^{8}$ km$^{2}$, lifetime: $\sim$8h, and magnetic flux: $\sim 10^{20}$Mx), microwave bright points, magnetic bright points, and He I $10830\AA$ dark points; (4)explosive phenomena, such as transition region explosive events and mini-filament eruptions; (5)blinkers; and (6)micro-flares and nano-flares (Golub et al. 1974; Wilhelm et al 2007; De Pontieu et al 2011; Zhang $\&$ Liu 2011; Longcope $\&$ Tarr 2015; Schmelz $\&$  Winebarger 2015; Tavabi et al. 2015). They are all distributed on the full solar disk.
A lot of case studies have demonstrated that these small-scale magnetic activity phenomena make a great contribution to coronal heating, and the corona is impulsively heated by them.  Ubiquitous small-scale  magnetic elements  are believed to contain the process of energy buildup and release in the solar corona in all probability, which are manifested everywhere on the solar disk as these small-scale magnetic activity phenomena (Zhang $\&$ Liu 2011; Testa et al 2014; Longcope $\&$ Tarr 2015; Schmelz $\&$  Winebarger 2015).

On the other hand for theory study, some models are proposed to address this issue, and they may be divided into two groups:  magnetohydrodynamic (MHD)  waves and magnetic reconnection energy releases (Alfven 1947; Parker 1972, 1988; Cranmer 2012; Arregui 2015;  Wilmot-Smith 2015).
Convective flows below the solar surface and/or the emergence of magnetic field cause a random shuffling and further twisting and braiding of the small-scale magnetic field lines. Magnetic field reconnection occurs at the braiding boundaries, creating a great deal of heat and plasma outflows, and the corona is heated by the cumulative effect of these small, localised activities (Narain $\&$ Ulmschneider 1996; Cranmer 2012; Longcope $\&$ Tarr 2015; Wilmot-Smith 2015). Observations demonstrate that both wave heating and reconnection heating work, but which one the main contributor for coronal heating is has been unknown now (Arregui 2015;  Wilmot-Smith 2015).

It has  been a bewildering mystery also up to now for the chromospheric heating (Narain $\&$ Ulmschneider 1996).  The standard model (VAL) of the quiet solar atmosphere shows that
temperature strangely increases from the top of the photosphere to the vicinity of the coronal base especially  with a rapid increase in the transient region (Vernazza et al 1981).
So far no compelling theories have been able to explain such a distribution of temperature (Narain $\&$ Ulmschneider 1996; Dunin-Barkovskaya $\&$ Somov 2016).
Therefore, heating is actually  a far-unaddressed issue at all layers of the upper solar atmosphere, although more attentions have been paid to coronal heating.

In this study, daily solar spectral irradiances at 985 spectral bands, which form at different layers of the solar atmosphere,
are utilized to investigate their (phase) relationship with solar magnetic activity, and then accidentally one statistical evidence arises for small-scale weak magnetic activity heating the upper solar atmosphere.

\section*{Observation and data reduction}
Solar spectral irradiance (SSI) at the spectral intervals 1-39 nm and 116-2416 nm, which are measured by the SORCE satellite during March 1 2003 to October 28 2017 can be available from the web site http://lasp.colorado.edu/home/sorce/data/. SSI is measured at 985 spectral bands which are all given as an attachment (Attachment I), and here as  samples  the time series of SSIs at 7  bands are shown in Figure 1.

It is not every day that
SSI is measured, and Figure 2 shows the number of measurement days for all spectral bands in the total of 5356 days. For those SSIs whose bands are shorter than 1600nm,
the number of measurement days is almost a constant, about 5000 days.
For the 985 bands, the maximum span of a band is 1 nm, and  thus wavelength (L) is used to replace ``spectral band" in the figure,  which is the middle value of a band.
Daily SSIs are used to investigate phase relation with daily sunspot number (SSN).
Daily SSN (version 2) at the same time interval is downloaded from http://sidc.oma.be/silso/, which is shown in Figure 3.
SSN itself is a kind of count of large-scale magnetic structures, but the time series of SSN is usually and here also used to reflect temporal occurrence frequency of large scale activity phenomena, for example,  flares, which are related to large-scale magnetic structures.
The Mount Wilson Sunspot Index (MWSI) and the Magnetic Plage Strength Index (MPSI) are calculated at the Mount Wilson observatory through daily magnetograms (Howard et al 1980).
Also used  are daily MWSI and MPSI during March 1 2003 to December 31 2012, which are available from http://obs.astro.ucla.edu/intro.html and shown in Figure 3. It is  not every day that MWSI and MPSI are observed, and they are obtained just in 2591  of the total 3594 days.
MWSI and MPSI  themselves  are a kind of count of strong magnetic fields  (mainly in sunspots) on the solar full disk and that of  weak magnetic fields  (mainly at outside of sunspots) (Howard et al 1980), correspondingly.
Here, the time series of the weak-magnetic-element index, MPSI is used to reflect temporal  occurrence frequency of the aforementioned small-scale magnetic activity phenomena, which are related with the weak magnetic elements, and the series of the strong-magnetic-element index, MWSI, the occurrence frequency of large-scale magnetic activity phenomena.

\section*{Data Analysis}
\subsection*{Phase Relation of SSI with SSN}
In order to study phase relationship of daily SSI with SSN, we perform a lagged cross-correlation analysis of SSN with  each  of the total 985 SSI time series.
First, letting two considered time series cover a same time interval (the two can be paired with each other every day), we calculate the cross-correlation coefficient (CC) of the two time series with no relative phase shifts. Next, one series is shifted by
one day with respect to the other, and
the unpaired endpoints of the two series are deleted. Then, we can obtain a new value
of CC, and it is the value at the relative shift of one day. Next again, the original series is shifted by
2 days with respect to the other original series, and the unpaired endpoints are deleted. Then, a new CC value can be obtained, which is the value at the relative shift of two days, and so on.
If no observation record for a time series is available on a certain day,
then no measurement value for that day is taken part in calculating CC.
As samples, Figure 4 shows the obtained CCs varying with relative shifts (called CC-phase lines below) for the seven sample  SSIs shown in Figure 1, where the
abscissa is shifts of SSN versus  SSI with backward
shifts given minus values, and the total 985 CC-phase lines are all given as an attachment (Attachment II).
CC-phase lines  peak around shifts being about 0 ($CC_{0}$) and $\pm 27$ days ($CC_{\pm 27}$), and thus solar rotation signal can be seen.
The local peak  values of each of the 985 CC-phase lines  around shifts being about 0 and $\pm 27$ days are given in Figure 5, which are the local maximum  when CCs around one peak are positive, or the local minimum  when CCs around one ``peak" (actually valley) are negative.
Their corresponding 12-point running averages are also given in the figure. As the figure shows, peak CCs may be divided into 7 parts.  For Part 1, $0.5 (nm) < L < 255.5 (nm)$,
$CC_{+27}$  is obviously larger than $CC_{-27}$, but less than $CC_{0}$. As an example, these three characteristic CC values for the spectral line whose spectral band is 8-9 nm are given in Table 1.
Following Li et al (2002), we test the statistical significance for difference of these  CCs by means of the Fisher translation method (Fisher 1915), which is given in the table. For Part 2, $256.5 (nm) < L < 288.5 (nm)$, $CC_{+27}$  is obviously significantly larger than $CC_{-27}$, and even larger than $CC_{0}$.
For Parts 1 and 2, CC-phase lines convexly peak around  $CC_{0}$.
For Part 3, $289.5 (nm) < L < 291.5 (nm)$, $CC_{+27}$ that is of statistical significance  is larger than  $CC_{0}$ that is statistically insignificant,  however, $CC_{+27}$ cannot be significantly larger than $CC_{0}$. From this part on, CC-phase lines start to be concaved around  $CC_{0}$.
For Part 4,  $292.5 (nm) < L < 802.42 (nm)$, $CC_{+27}$  is significantly larger than $CC_{0}$. Both $CC_{+27}$ and $CC_{0}$ are positive values of statistical significance.
For Part 5,  $806.05 (nm) < L < 876.5 (nm)$, both $CC_{0}$ and $CC_{+27}$ change from positive values of statistical insignificance to  negative values of statistical insignificance.
For Part 6,  $880.71 (nm) < L < 1598.95 (nm)$, $CC_{0}$ is a negative value of statistical significance.
For Part 7,  $1601.18 (nm) < L < 2412.34 (nm)$, both $CC_{+27}$ and $CC_{0}$ disorderly vary with wavelength, and this part is no longer taken into account below.
For one of the 7 parts, one spectral line is chosen as a sample, which is shown in Figure 1, and
its three characteristic CCs ($CC_{0}$ and $CC_{\pm 27}$) and statistical significance test for their difference are given in Table 1.

\begin{center}
\begin{table}
\begin{center}
 \caption{Statistical significance for difference of characteristic CCs of SSI vs SSN}
 \begin{tabular}{cccccc}
 \hline
 Spectral band & $CC_{0}$ &  $CC_{+27}$ & $CC_{-27}$ &   Probability         &    Probability       \\
               &            &           &            & $CC_{+27}$ vs $CC_{0}$   &  $CC_{+27}$ vs $CC_{-27}$ \\
\hline
8-9 nm         &  0.8664    & 0.7827    & 0.7311     &      $> 99.9\%$   &    $> 99.9\%$     \\
266-267 nm     &  0.5801    & 0.6195    & 0.5583     &   $97\%$                 &    $> 99.9\%$     \\
289-290 nm     &  0.0811    & 0.1144    & 0.0882     &    $76\%$                &                          \\
431.47 nm      &  0.3966    & 0.5061    & 0.4377     &      $> 99.9\%$   &                          \\
843.96 nm      &  0.020     & 0.1016    & 0.0438     &   $99.5\%$               &                          \\
1040.57 nm     &  -0.2434   & -0.2021   & -0.2185    &   $88\%$                 &                          \\
1616.86 nm     &  -0.03930  & -0.0518   & -0.0325    &                          &                          \\
  \hline
 \end{tabular}
\end{center}
 \end{table}
\end{center}

Generally, $CC_{+27}$ is significantly larger than $CC_{-27}$ for the SSI whose wavelength is shorter than $\sim 800$ nm. And further, $CC_{+27}$ is even significantly larger than $CC_{0}$ for the SSI whose wavelength is shorter than $\sim 800$ nm but longer than $\sim 300$ nm, implying that SSN should lead these SSIs by about a solar rotation. Strong magnetic field in sunspot regions generally becomes weak magnetic field after a solar rotation, and thus this phase lead means that the SSI should be more related to weak magnetic field activity than strong magnetic field activity.

\subsection*{Phase Relation of SSI with MWSI and MPSI}
Similarly, we perform a lagged cross-correlation analysis of the total 985 SSI time series respectively with MPSI and MWSI.
As samples, Figure 6 shows the obtained CCs varying with relative phase shifts  for the first six sample  SSIs, where the
abscissa is shifts of MPSI (or MWSI) versus  SSI with backward shifts given minus values.
The solar rotation signal can be seen when MPSI/MWSI significantly leads SSI by about a solar rotation.
Table 2 shows $CC_{0}$s and $CC_{+27}$s of the first six sampled lines. $CC_{0}$ is even up to 0.9366 for the first sample line, indicating that the corresponding SSI should be closely related to weak magnetic field activity.
Similarly, the local peak  values of a CC-phase line  around shifts being about 0 and $\pm 27$ days are given in Figure 7 for SSI vs MPSI and in Figure 8 for SSI vs MWSI,  and
their 12-point averages are also given in the corresponding figures.
Statistical significance test for difference of three characteristic CCs  may be referred to Table 2.
Peak CCs in Figures 7 and 8 can be divided into 7 same parts as done in Figure 5.

\begin{center}
\begin{table}
\begin{center}
 \caption{Statistical significance for difference of characteristic CCs of SSI with MPSI and MWSI}
 \begin{tabular}{cccccccc}
 \hline
               & \multicolumn{3}{c}{SSI vs MPSI}       & \multicolumn{3}{c}{SSI vs MWSI}        &     Probability  \\
 Spectral band & $CC_{0}$ &  $CC_{+27}$ &   Probability  & $CC_{0}$ &  $CC_{+27}$ &    Probability  & for  these two $CC_{0}$s  \\
\hline
8-9 nm         &  0.9366  &  0.8660  &    $> 99.9\%$ &  0.7289  &  0.6166  &    $> 99.9\%$ &    $> 99.9\%$ \\
266-267 nm     &  0.7381  &  0.7445  &  $72\%$              &  0.4478  &  0.5230  &    $> 99.9\%$ &    $> 99.9\%$ \\
289-290 nm     &  0.2759  &  0.3022  &  $53\%$              &  0.1677  &  0.2512  &  $97\%$     &  $99\%$ \\
431.47 nm      &  0.5151  &  0.6184  &    $> 99.9\%$ &  0.0820  &  0.4317  &    $> 99.9\%$ &    $> 99.9\%$ \\
843.96 nm      &  0.1805  &  0.2359  &  $85\%$              & -0.1981  &  0.1219  &    $> 99.9\%$ &    $> 99.9\%$ \\
1040.57 nm     & -0.5324  & -0.4600  &  $98\%$              & -0.6541  & -0.3145  &    $> 99.9\%$ &    $> 99.9\%$ \\
  \hline
 \end{tabular}
\end{center}
 \end{table}
\end{center}

We choose those SSNs which are recorded on the same days as MPSI being observed (the two are simultaneously observed), which are here called the chosen SSN, and then perform a lagged cross-correlation analysis of the chosen SSN  respectively with the 985 SSI time series. Similarly, Figure 9 shows the local peak  values of a CC-phase line  around shifts being about 0 and $\pm 27$ days. As Figures 7 to 9 display, generally $CC_{+27}$ is significantly larger than $CC_{-27}$ for SSI whose wavelength is shorter than $\sim 800$ nm, implying that weak magnetic activity should influence SSI after a solar rotation. $CC_{+27}$ of both SSN and MWSI respectively with SSI is  even significantly larger than the corresponding $CC_{0}$ for  SSI whose wavelength is shorter than $\sim 800$ nm but longer than $\sim 300$ nm, implying that strong magnetic field activity (SSN and MWSI) should lead SSI by about a rotation.
Three $CC_{0}$ lines shown in Figures 7 to 9 are put together in Figure 10, and then this figure clearly displays that  SSI whose wavelength is shorter than $\sim 800$ nm (Parts 1 to 4) should be much more related to MPSI than to MWSI, and that  SSI whose wavelength is longer than $\sim 880$ nm and shorter than $\sim 1600$ (Part 6) should generally be more related to MWSI than to MPSI.  $CC_{0}$  of MPSI  with SSI at X-rays and  far ultraviolet is
obviously larger than that at visible light  band, implying that  the relationship of small-scale weak magnetic activity with SSI is much more intimate
at X-rays and  far ultraviolet than that at visible light  band.
$CC_{0}$ for SSI vs the chosen SSN is located between $CC_{0}$ for SSI vs MPSI  and $CC_{0}$ for SSI vs MWSI, implying that the magnetic fields of sunspots which are all counted  to SSN  should be counted partly  into MWSI and partly into MPSI.

Comparison of Figures 5 with 9 shows that, data missing  should change values of CC, but the relative trend of  $CC_{0}$ and  $CC_{+27}$ still exists, which is slightly influenced.

\section*{Conclusions and Discussion}
X-rays and far ultraviolet spectra form in the transition region and  corona, middle and near ultraviolet spectra form over or in the top chromosphere, SSIs at visible-light wavelengths, $400 \sim 800$nm form in the chromosphere or in the upper photosphere, and SSIs  at infrared wavelengths, $900 \sim 1600$nm mainly form in the low photosphere (Vernazza et al 1981; Ding $\&$ Fang 1989; Harder et al 2009; Meftah et al. 2018). Therefore, Figures 5 and 10 indicates that in the low photosphere SSI is negatively correlated with SSN  and more related to strong magnetic activity than to  weak magnetic activity, but in and above the top photosphere, SSI is positively correlated with SSN, and significantly more related to weak magnetic activity than to strong magnetic activity.
Correspondingly, as the VAL atmosphere model (Vernazza et al 1981) shows, temperature decreases towards the outside  in the low photosphere, but abnormally increases from the top photosphere up towards the corona.
Therefore, the layer at which temperature  decreases (the unheated region) is found to correspond to the layer at which SSI is in anti-phase with SSN  and more related to strong magnetic activity than to  weak magnetic activity, and the layer at which temperature   abnormally increases (the heated region) is found to be the layer at which SSI is in phase with SSN  and more related to weak  magnetic activity.  The heated region accurately corresponds to the region where weak field activity has a more obvious impact on SSI, and outside the region, strong magnetic activity does instead.
Long-term variation of ``energy" in the entire heated region, from  the top photosphere, chromosphere to  corona  dances in step with  weak magnetic activity.
Long-term fluctuation of the SSI forming at the bottom of the photosphere, which reflects long-term variation of ``energy" leaked from the solar interior, completely differ from that of the SSI forming in the heated region. Therefore,
it should naturally be weak magnetic activity that causes the abnormal temperature distribution.
At the base of the heated region, namely in the top photosphere and chromosphere, strong magnetic activity is found to lead SSI by about one solar rotation.
Solar strong magnetic structures (sunspots) are first observed when they appear on the solar disk, and their evolutional
components should be observed after  rotating to appear again on the solar visible disk as the disintegrated components (small-scale weak magnetic elements).
Because sunspots are darker than the background photosphere and  chromosphere and become weak magnetic elements after a solar rotation, $CC_{+27}$ of both SSI and MWSI (strong magnetic field index) vs SSI is obviously larger than the corresponding $CC_{0}$ in the top photosphere and  chromosphere, displaying that strong magnetic activity (both the time series of magnetic structures and the temporal series of active/eruptive phenomena related to them) should lead SSI by about a solar rotation, and such the phase lead also implies weak magnetic activity  heating.
These results are illustrated in Figure 11.
Further, more reconnection events of small-scale magnetic activity are observed at higher atmosphere layer, the transition region and corona, and thus temperature is  higher at the higher layer. Correspondingly, the relation of small-scale magnetic activity (MPSI) with the SSIs at short wavelengths, X-rays and  far ultraviolet,
forming at the higher  layer  is obviously closer than that of MPSI with  the SSIs at long wavelengths (visible light  band),
forming  in the top photosphere and chromosphere, and the time series of the SSIs at the short wavelengths more violently fluctuate  (see SSI time series given in the Figure 1and Attachment I), namely, the SSIs at the short wavelengths more obviously respond to weak magnetic activity. All findings  point the finger of heating the chromosphere, transition region and corona  firmly at weak magnetic activity.

Recently through analyzing observational data of the quiet Sun by the Interface Region Imaging Spectrograph,
Tavabi (2018) found a strong relationship among the network bright points at all layers of the solar atmosphere and suggested that magnetic-field concentrations in the network rosettes should helpfully couple between the solar inner and outer atmosphere. Thus, our statistical result is supported by observations, and synchronously heating all layers of high solar atmosphere (from the upper photosphere to the corona) is feasible and actually so indeed.

The heating of waves emerging from the solar interior should not be the main heating mechanism of  high solar atmosphere  due to the following reasons.
(1) The physical situation  is very different from one another in the upper photosphere,  chromosphere, transition region, and  corona, thus synchronously heating these layers seems unlikely through wave dissipation. And
(2) it is hard to explain that those SSIs which form at these layers are all in phase with SSN and more related with weak magnetic activity (MPSI), and especially hard to explain the phase lead of strong magnetic activity with respect to SSI  by wave  heating.
Of course, contributions of waves to heating  high solar atmosphere have been observed.
After excluding the wave-heating from  the main heating mechanism, the remaining reconnection-heating mechanism  can indeed explain well the strong relation of those SSIs with weak magnetic activity and such the phase lead.

As claimed at the first section, high-resolution observations point the finger of coronal heating to small-scale weak magnetic activity.
Especially, magnetic reconnection was found to occur at a much small spatial scale throughout the solar chromosphere by Shibata et al  (2007),
and they believe that the heating of the chromosphere and  corona should be connected to the small-scale ubiquitous reconnection.
Up to now, high-resolution observations have given  evidences for weak magnetic activity heating high solar atmosphere
just through local heating channel, but how the heated atmosphere is globally related to weak magnetic activity (ubiquitous small-scale active events) has not been investigated, and it is hard to address this issue.
Here we do not yet directly answer this question, but the comparison of the temporal variation (SSIs) of the heated atmosphere is carried out with weak magnetic activity, which is represented by weak magnetic index (MPSI). Our findings put it to the proof from the final effect of global heating, and exclude  other possible mechanisms of heating  known at present.
Our statistical evidence is compelling, and even irrefutable when observational evidences are combined.

Sunspots present themselves as dark constructs in the photosphere and  low and middle chromosphere, but bright in the top chromosphere and  transition region (Ding $\&$ Fang 1989).
Therefore, the aforementioned CC-phase lines of SSN (or MWSI) vs SSI  convexly peak around  $CC_{0}$ for SSIs whose wavelengths are shorter than about 290 nm (appearance of sunspots on the visible disk as bright  constructs should lead to increasing SSI),  but are concaved for SSIs whose wavelengths are longer than about 300 nm (sunspots  present as dark  constructs  to decrease SSI).
Some of the SSIs  at the infrared wavelengths, $1600 \sim 2400$nm (Part 7) may form in the low photosphere, some form in or over the top  photosphere, and some  even possibly come from the both, thus some SSIs at this spectral interval are in phase with SSN, but some are not. Therefore, SSIs at this spectral interval disorderly vary with wavelength.

Finally, it should be emphasized that some magnetic indexes (SSN,  Mg II index, and so on) are used as ``proxy" indexes in the research field of SSI/TSI (total solar irradiance) reconstruction (Lean 2000; Frohlich 2006; Steiner 2007; Fontenla et al. 2011; Ermolli et al. 2014, Yeo et al 2014; Dudok de Wit et al 2018; and references therein),  which act for appearance of magnetic structures on the solar disk, but here time series of  magnetic indexes are used to reflect temporal  occurrence frequency of
magnetic activity phenomena (events), which are related to the corresponding magnetic structures of the indexes.
The temporal variation of the heated high atmosphere is reflected by the SSIs which form there, and
the temporal variation of occurrence frequency of small-scale weak magnetic activity phenomena is reflected here by MPSI.
On the one hand, the special (phase) relation of SSIs and MPSI, which clearly differs from that of  SSIs and MWSI/SSN,
deny  the main heating mechanism of wave heating and remove major contributions of strong magnetic activity to heating high atmosphere.
On the other hand, the primary reason why
those SSIs forming at high atmosphere layers  vary in the long term should be the heating of small-scale weak magnetic activity phenomena.

{\bf acknowledgments:}
The authors thank the anonymous referee for careful reading of the manuscript and constructive
comments which improved the original version of the manuscript.
We would like to thank Profs. Jingxiu Wang and Mingde Ding for helpful discussion and useful suggestions.
The SORCE Solar Spectral Irradiance (SSI) composite data product is
constructed using measurements from the XPS, SOLSTICE, and SIM instruments, which are combined into merged daily solar spectra over the
spectral intervals (1-39 nm and 116-2416 nm).
Sunspot data come from the World Data Center SILSO, Royal Observatory of Belgium, Brussels.
This study includes data (MPSI and MWSI) from the synoptic program at the150-Foot Solar Tower of the Mt. Wilson Observatory. The Mt. Wilson 150-Foot Solar Tower is operated by UCLA, with funding from NASA, ONR and NSF, under agreement with the Mt. Wilson Institute.
All data used here are  downloaded from web sites.
The authors would like to express their deep thanks to the staffs of
these web sites. This work is supported by the
National Natural Science Foundation of China (11573065  and
11633008)  and the Chinese Academy of Sciences.

\clearpage

\begin{figure*}
\begin{center}
\centerline{\includegraphics[width=.8\textwidth]{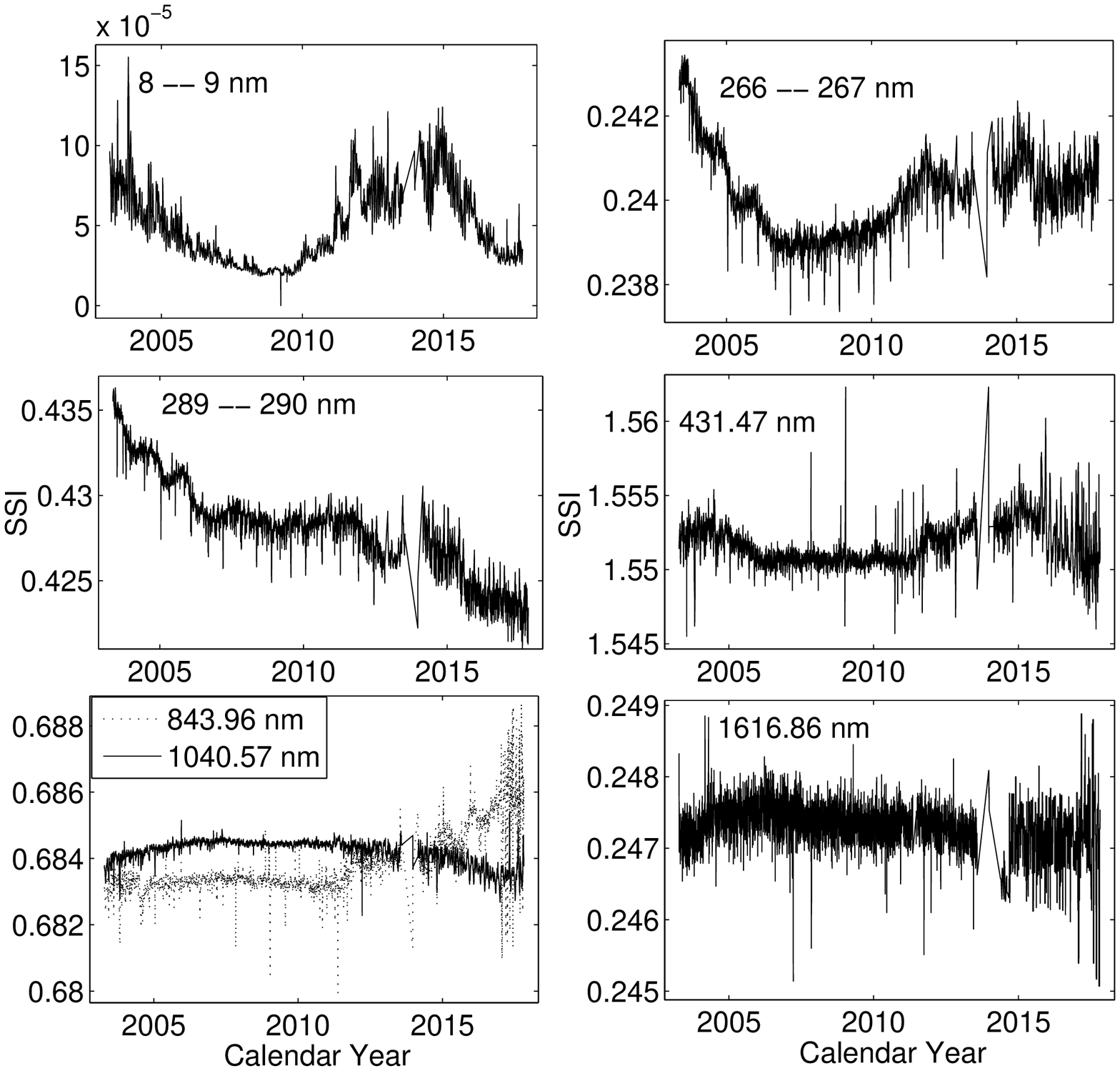}}
\caption{Time series of SSI at 7 spectral bands. The band of a SSI is given at the upper and left corner of a panel.
}\label{}
\end{center}
\end{figure*}

\begin{figure*}
\begin{center}
\centerline{\includegraphics[width=.8\textwidth]{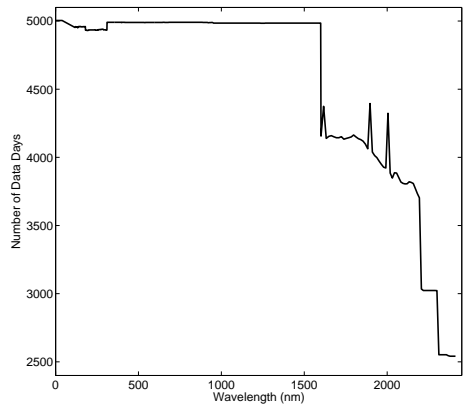}}
\caption{The number of data days for SSI at a band observed during March 1 2003 to Oct. 28 2017.
}\label{}
\end{center}
\end{figure*}

\begin{figure*}
\begin{center}
\centerline{\includegraphics[width=.8\textwidth]{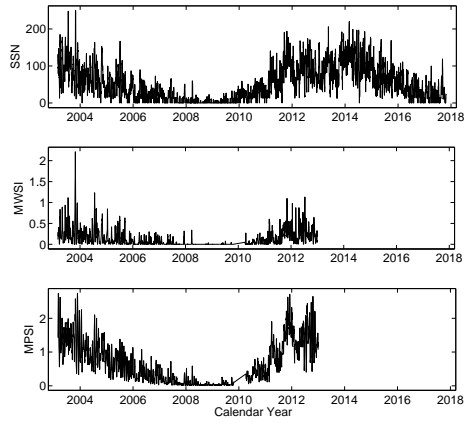}}
\caption{Top panel: daily SSN from March 1 2003 to Oct. 28 2017.
Middle panel: daily MWSI from March 1 2003 to Dec. 31 2012.
Bottom panel: daily MPSI from March 1 2003 to Oct. 28 2017.
}\label{}
\end{center}
\end{figure*}

\begin{figure*}
\begin{center}
\centerline{\includegraphics[width=.8\textwidth]{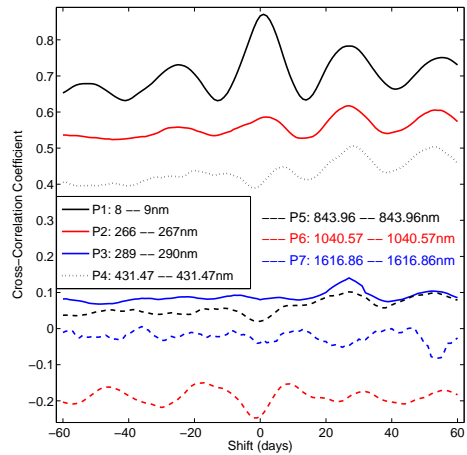}}
\caption{Cross-correlation coefficient of daily sunspot number respectively with the 7 sampled SSI lines shown in Figure 1,
varying with their relative phase shifts  with backward shifts given minus values.
}\label{}
\end{center}
\end{figure*}

\begin{figure*}
\begin{center}
\centerline{\includegraphics[width=.8\textwidth]{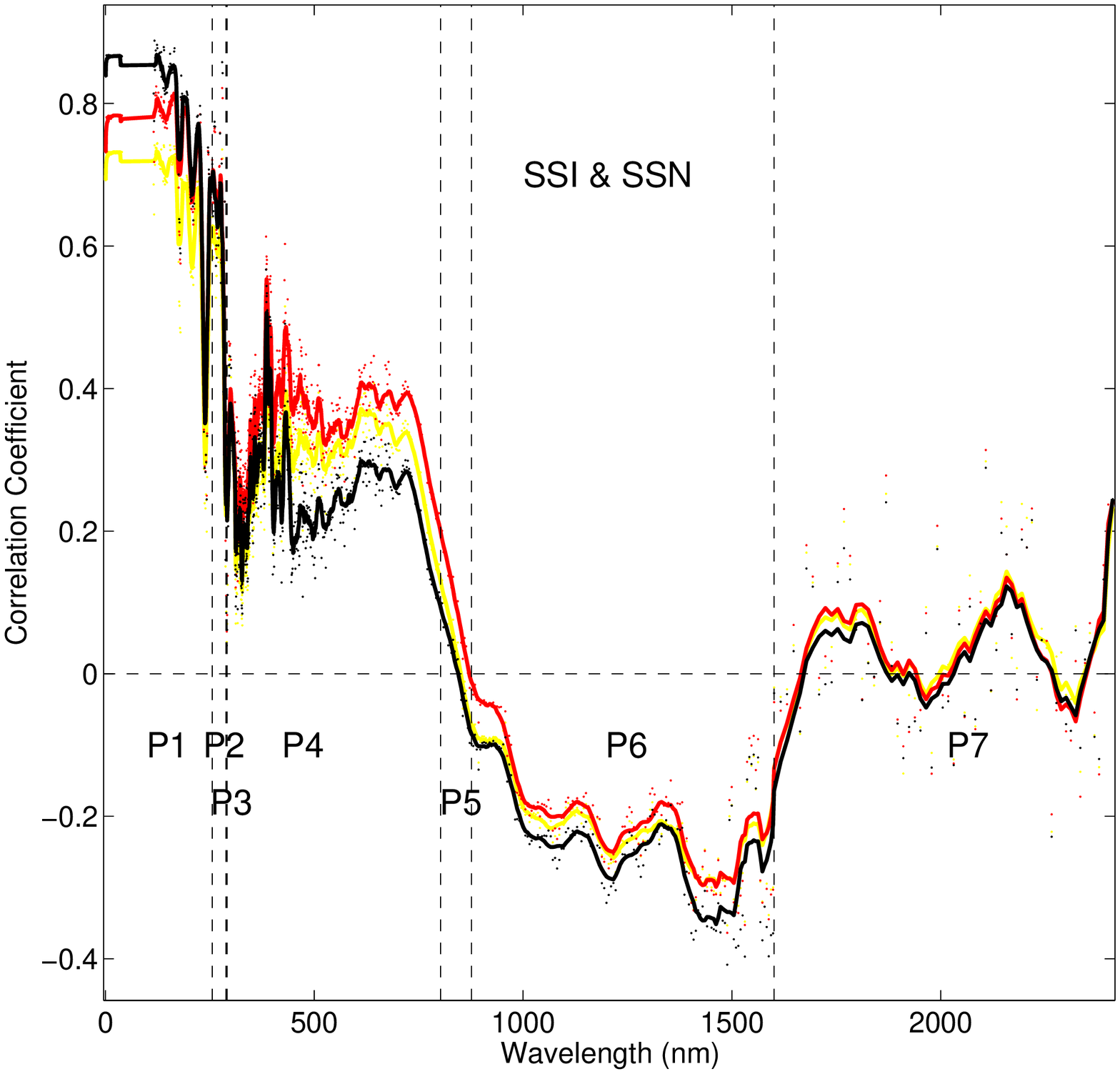}}
\caption{Local peak  value of a cross-correlation coefficient line (SSN vs SSI)  around shifts being about 0 (black dots), -27 (yellow dots) and 27 days (red dots).
The solid lines are correspondingly 12-point smoothing averages.
}\label{}
\end{center}
\end{figure*}

\begin{figure*}
\begin{center}
\centerline{\includegraphics[width=.8\textwidth]{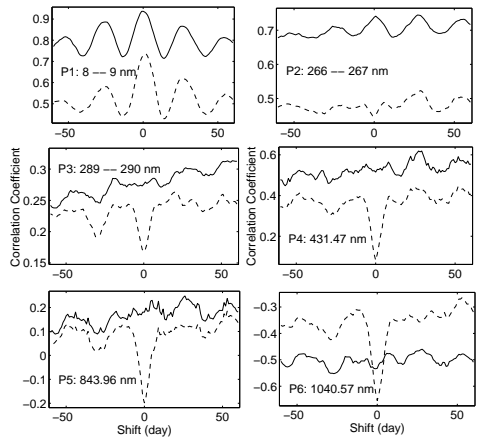}}
\caption{Cross-correlation coefficient of the first 6 sampled SSI lines shown in Figure 1 respectively with  MPSI (solid line) and MWSI (dashed line)   ,
varying with their relative phase shifts  with backward shifts given positive values.
}\label{}
\end{center}
\end{figure*}

\begin{figure*}
\begin{center}
\centerline{\includegraphics[width=.8\textwidth]{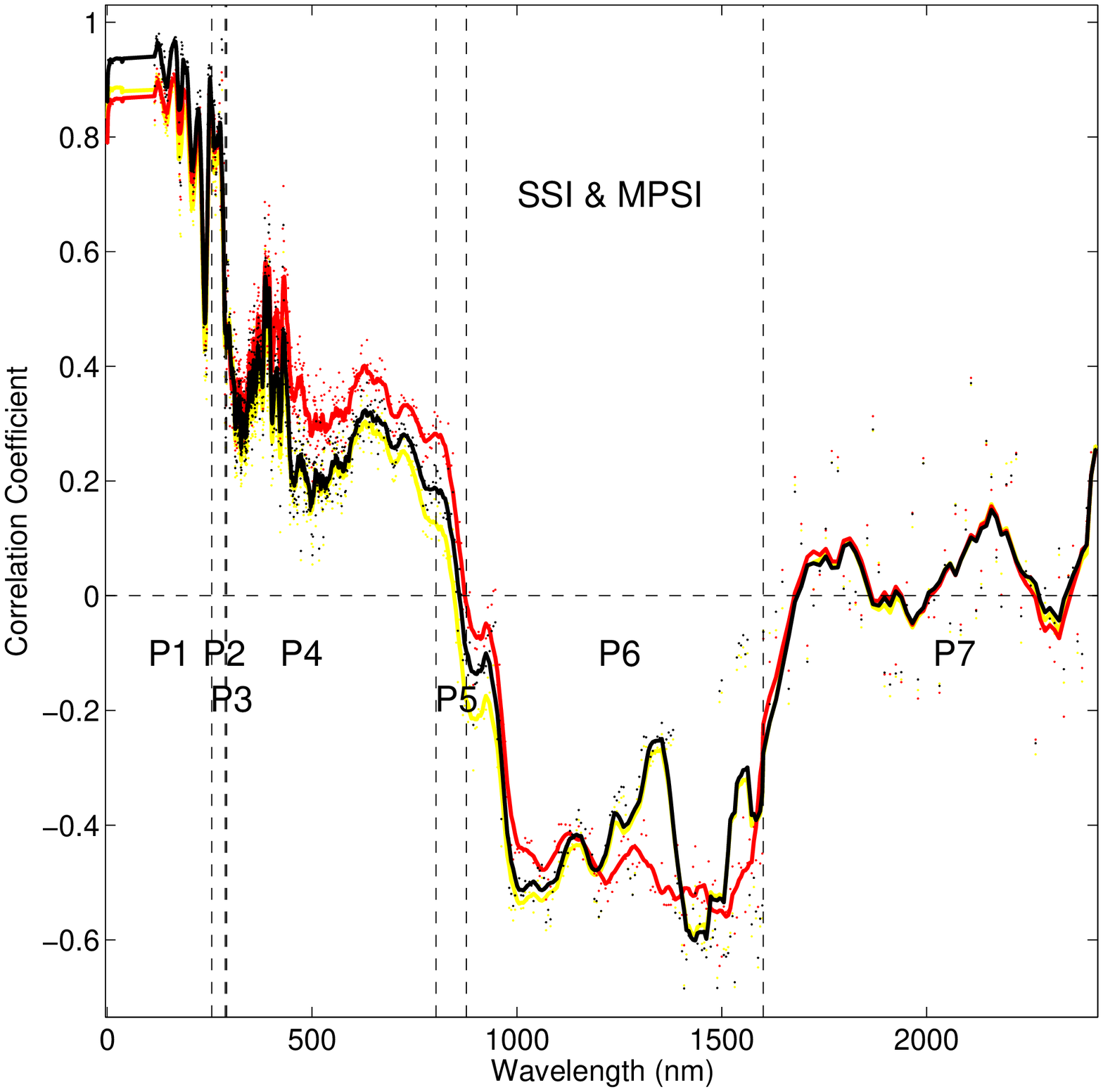}}
\caption{Local peak  value of a cross-correlation coefficient line (MPSI vs SSI)  around shifts being about 0 (black dots), -27 (yellow dots) and 27 days (red dots).
The solid lines are correspondingly 12-point smoothing averages.
}\label{}
\end{center}
\end{figure*}

\begin{figure*}
\begin{center}
\centerline{\includegraphics[width=.8\textwidth]{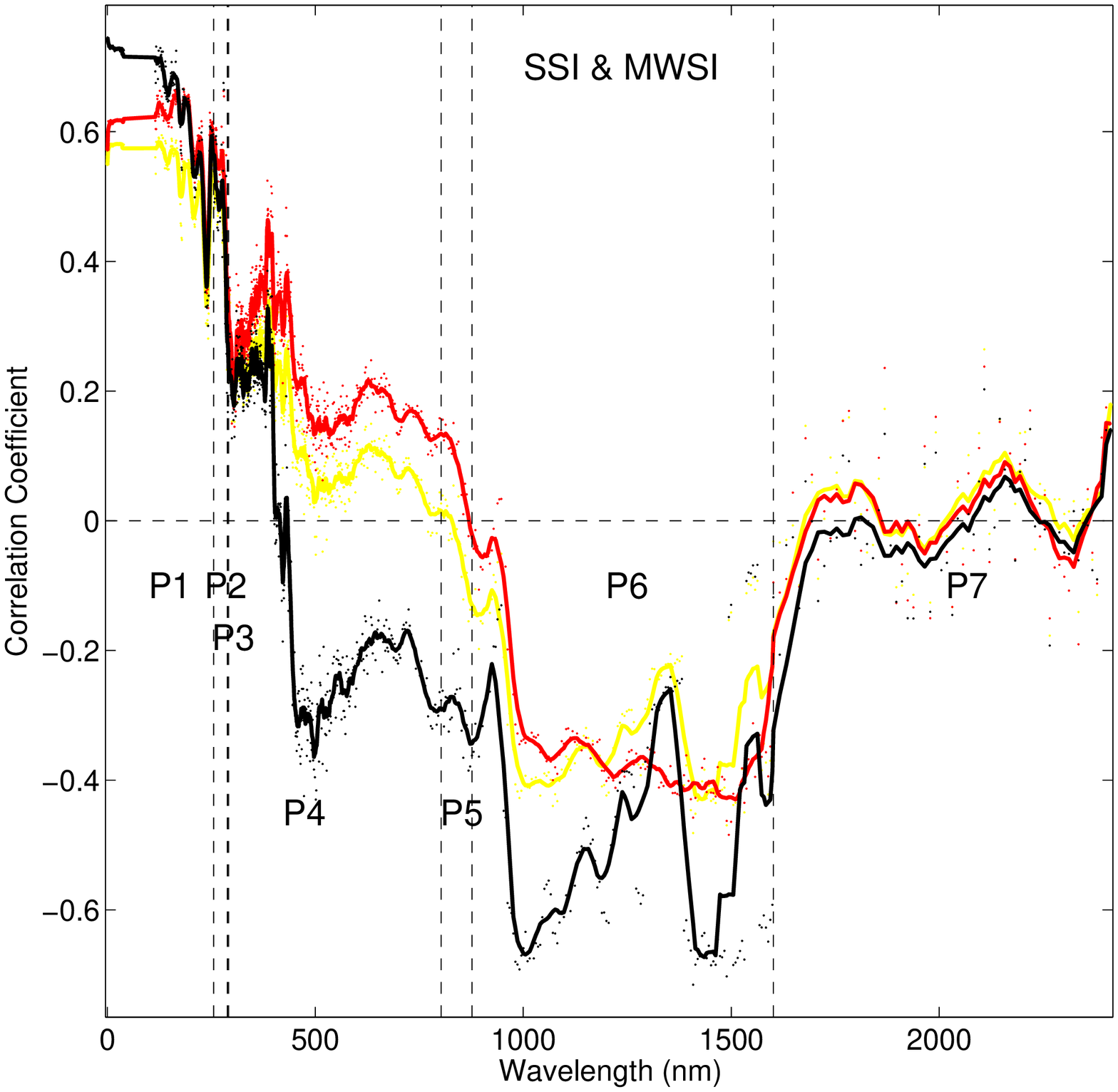}}
\caption{Local peak  value of a cross-correlation coefficient line (MWSI vs SSI)  around shifts being about 0 (black dots), -27 (yellow dots) and 27 days (red dots).
The solid lines are correspondingly 12-point smoothing averages.
}\label{}
\end{center}
\end{figure*}

\begin{figure*}
\begin{center}
\centerline{\includegraphics[width=.8\textwidth]{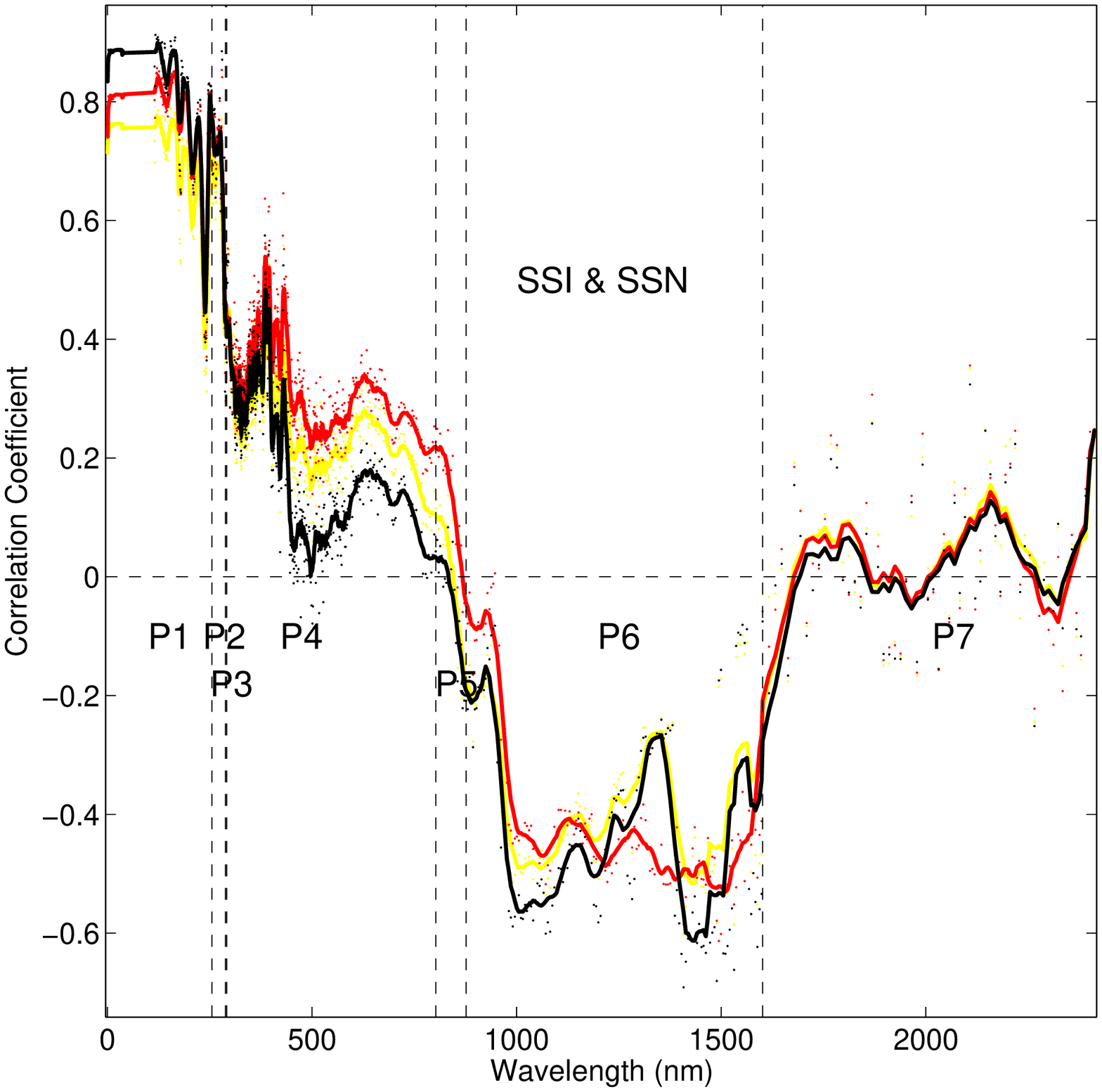}}
\caption{Local peak  value of a cross-correlation coefficient line (the chosen SSN vs SSI)  around shifts being about 0 (black dots), -27 (yellow dots) and 27 days (red dots).
The solid lines are correspondingly 12-point smoothing averages.
}\label{}
\end{center}
\end{figure*}

\begin{figure*}
\begin{center}
\centerline{\includegraphics[width=.8\textwidth]{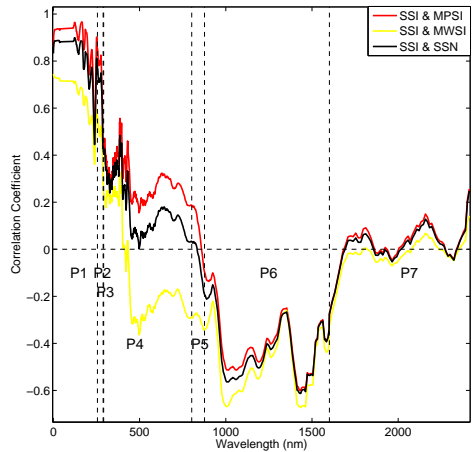}}
\caption{Comparison of three $CC_{0}$: $CC_{0}$ for SSI vs MPSI (red line), $CC_{0}$ for SSI vs MWSI (yellow line), and $CC_{0}$ for SSI vs the chosen SSN (black line).
}\label{}
\end{center}
\end{figure*}

\begin{figure*}
\begin{center}
\centerline{\includegraphics[width=.8\textwidth]{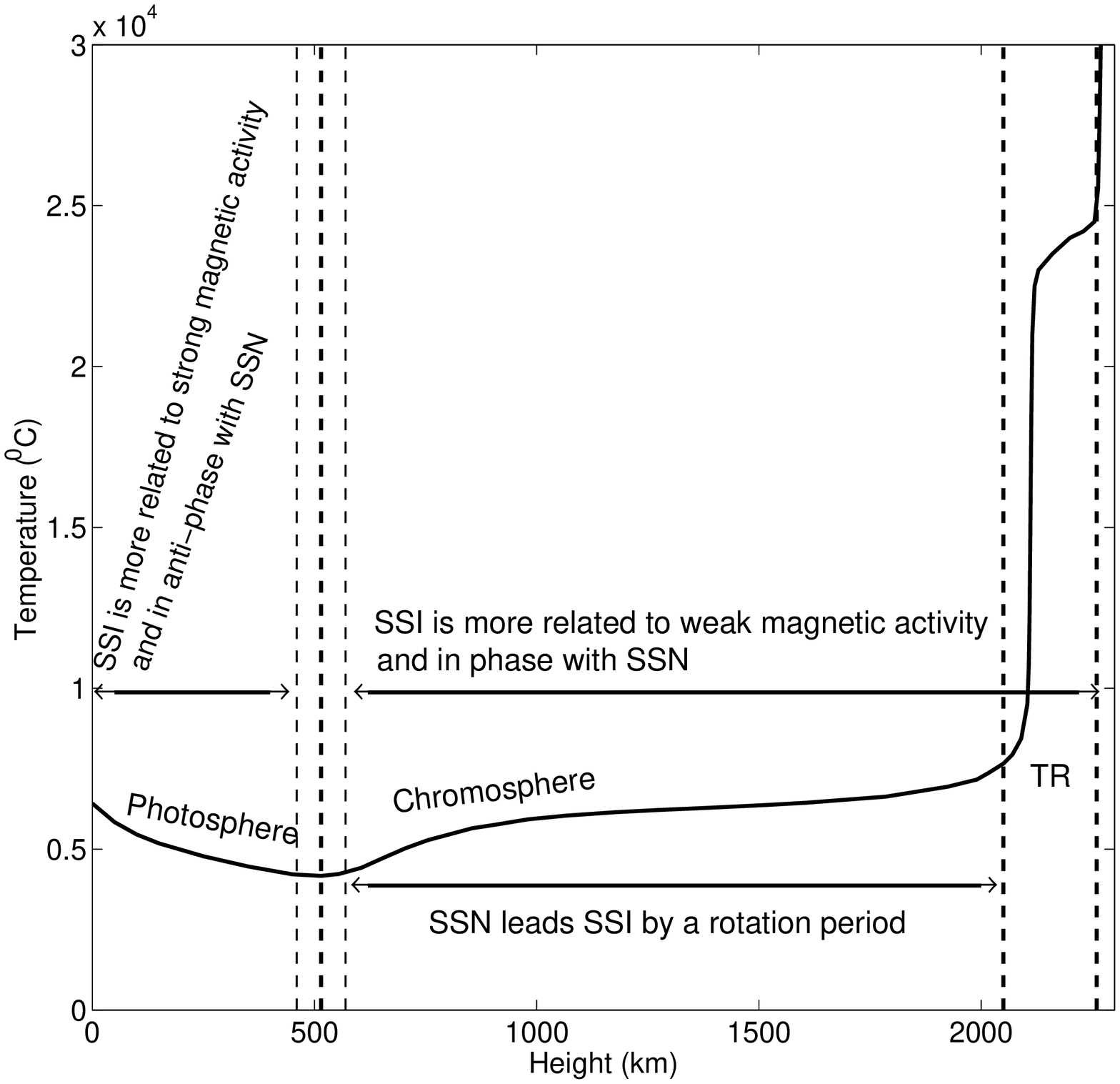}}
\caption{Diagrammatic sketch of relation  of solar magnetic activity with SSI at different solar atmosphere layers. Based on the sketchy height distribution of temperature (the thick solid line), which is given by the VAL model, the solar atmosphere is divided into the photosphere, chromosphere, transition region (TR), and corona,  which are separated by  vertical thick dashed lines.
}\label{}
\end{center}
\end{figure*}


\begin{thebibliography}{}
\bibitem[ ]{}Alfven, H.  1947, MNRAS,  107, 211
\bibitem[ ]{}Arregui, I.  2015,  Philosophical Transactions of the Royal Society A: Mathematical, Physical and Engineering Sciences,   373, 20140261
\bibitem[ ]{}Cranmer, S. R.  2012, SSRv,   172, 145
\bibitem[ ]{}De Moortel, I., $\&$ Browning, P. 2015, Philosophical Transactions of the Royal Society A: Mathematical, Physical and Engineering Sciences,  373,  20140269
\bibitem[ ]{}De Pontieu, B., McLntosh, S. W., Carlsson, M. et al.  2011, Science,  331, 55
\bibitem[ ]{}Ding, M. D., $\&$ Fang, C. 1989, A$\&$A, 225, 204
\bibitem[ ]{}Dudok de Wit, T., Kopp, G., Shapiro, A., Witzke, V., $\&$ Kretzschmar, M. 2018, ApJ, 853, 197
\bibitem[ ]{}Dunin-Barkovskaya, O. V., $\&$ Somov, B. V. 2016, AstL,   42, 825
\bibitem[ ]{}Edlen, B. 1945, MNRAS, 105, 323
\bibitem[ ]{}Ermolli, I., Matthes, K., Dudok de Wit, T., Krivova, N. A., Tourpali, K., Weber, M., et al. 2013, Atmospheric Chemistry and Physics, 13, 3945
\bibitem[ ]{}Fisher, R. A. 1915, Biometrika,  10, 507
\bibitem[ ]{}Fontenla, J. M., Harder, J., Livingston, W., Snow, M., $\&$ Woods, T.  2011, JGRD,116, D20108
\bibitem[ ]{}Frohlich, C. 2006, SSRv, 125, 53
\bibitem[ ]{}Golub, L., Krieger, A. S., Silk, J. K., Timothy, A. F., $\&$ Vaiana, G. S. 1974, ApJ, 189, L93
\bibitem[ ]{}Harder, J. M.,   Fontenla, J. M., Pilewskie, P.,  Richard, E. C., $\&$ Woods, T. N.  2009, Geophysical Research Letters,  36, L07801
\bibitem[ ]{}Howard, R., Boyden, J. E., $\&$ LaBonte, B. J. 1980, SoPh,   66, 167
\bibitem[ ]{}Klimchuk, J. A. 2006, SoPh,  234, 41
\bibitem[ ]{}Klimchuk, J. A.  2015, Philosophical Transactions of the Royal Society A: Mathematical, Physical and Engineering Sciences,  373, 20140256
\bibitem[ ]{}Lean, J. L.  2000, SSRv, 94, 39
\bibitem[ ]{}Li, K. J., Irie, M., Wang, J. X., Xiong, S. Y., Yun, H., Liang, H. F., Zhan, L. S., $\&$ Zhao, H. Z. 2002, PASJ,   54, 787
\bibitem[ ]{}Longcope, D. W., $\&$ Tarr, L. A.  2015, Philosophical Transactions of the Royal Society A: Mathematical, Physical and Engineering Sciences,  373, 20140263
\bibitem[ ]{}Meftah, M., Dame, L., Bolsee, D.,  Hauchecorne, A.,  Pereira, N., Sluse, D.,  etal.  2018, A$\&$A, 611, A1
\bibitem[ ]{}Narain, U., $\&$ Ulmschneider, P.  1996, SSRv, 75, 453
\bibitem[ ]{}Parker, E. N. 1972, ApJ, 174, 499
\bibitem[ ]{}Parker, E. N. 1988, ApJ, 330, 474
\bibitem[ ]{}Parnell, C. E., $\&$ De Moortel, I. 2012, Philosophical Transactions of the Royal Society A: Mathematical, Physical and Engineering Sciences,  370, 3217
\bibitem[ ]{}Schmelz, J. T.,  $\&$  Winebarger, A. R. 2015, Philosophical Transactions of the Royal Society A: Mathematical, Physical and Engineering Sciences,  373, 20140257
\bibitem[ ]{}Shibata, K., Nakamura, T., Matsumoto, T. et al. 2007, Science,  318, 1591
\bibitem[ ]{}Steiner, O. 2007, AIP Conference Proceedings, 919, 74
\bibitem[ ]{}Tavabi, E. 2018, MNRAS, 476, 868
\bibitem[ ]{}Tavabi, E., Koutchmy, S., $\&$ Golub, L. 2015, SoPh, 290, 2871
\bibitem[ ]{}Testa, P., De Pontieu, B., Allred,J., Carlsson,M., Reale,F., Daw, A., Hansteen,V.,  Martinez-Sykora,J., Liu, W.  et al. 2014, Science,  346, 1255724
\bibitem[ ]{}Vernazza, J. E., Avrett, E. H., $\&$ Loeser, R.  1981, ApJS 45,  635
\bibitem[ ]{} Wilhelm, K., Marsch, E.,  Dwivedi, B. N., $\&$ Feldman, U. 2007, SSRv,  133, 103
\bibitem[ ]{}Wilmot-Smith, A. L.  2015, Philosophical Transactions of the Royal Society A: Mathematical, Physical and Engineering Sciences, 373, 20140265
\bibitem[ ]{}Yeo, K. L., Krivova, N. A., $\&$ Solanki, S. K. 2014, SSRv, 186, 137
\bibitem[ ]{}Zhang, J.  $\&$ Liu, Y. 2011, ApJL, 741, L7(5pp)
\bibitem[ ]{}Zirin, H. 1988, Astrophysics of the Sun, Cambridge Univ. Press, Cambridge¡¡
\end{thebibliography}
\end{document}